\documentclass[a4paper,11pt]{article}

\usepackage{graphicx}
\usepackage{amsmath}
\usepackage{amsfonts}
\usepackage{amssymb}
\usepackage{units}
\usepackage{epsf}
\usepackage{bbm}   % Blackboard-Math-Schriften, die die 1 enthalten
\newcommand{\be}{\begin{equation}}
\newcommand{\ee}{\end{equation}}
\newcommand{\ba}{\begin{align}}
\newcommand{\ea}{\end{align}}
\newcommand{\tr}{\mathrm {tr}}

\newcommand{\nn}{\nonumber \\}

\newcommand{\braket}[2]{\ensuremath{\langle \, #1 \, |\, #2 \, \rangle}}
\newcommand{\bra}[1]{\ensuremath{\langle \, #1 \, |}}
\newcommand{\ket}[1]{\ensuremath{| \, #1 \, \rangle }}

\newcommand{\cI}{\ensuremath{\mathcal{I}}}

\newcommand{\1}{\ensuremath{\mathbbm{1}}}

\begin{document}
\title{CP Violation and Arrows of Time:\\Evolution of a Neutral $K$ or $B$ Meson from an
 Incoherent to a Coherent State}
\author{Ch. Berger
%\thanks{\noindent I. Physikalisches Institut der RWTH, Aachen, Germany}
\\{\small I. Physikalisches Institut der RWTH, Aachen, Germany} \and
L. M. Sehgal\thanks{e-mail: sehgal@physik.rwth-aachen.de}
\\{\small Institut f{\"u}r Theoretische Physik (E) der  RWTH, Aachen, Germany}}

\maketitle

%\begin{center}{\Large{\bf Abstract}} \end{center}
\begin{abstract}
We study the evolution of a neutral $K$ meson prepared as an
incoherent equal mixture of $K^0$ and $\bar{K^0}$. Denoting the
density matrix by $\rho(t) = \frac{1}{2} N(t) \left[\1 +
\vec{\zeta}(t) \cdot \vec{\sigma} \right] $, the norm of the state
$N(t)$ is found to decrease monotonically from one to zero, while
the magnitude of the Stokes vector $|\vec{\zeta}(t)|$ increases
monotonically from zero to one. This property qualifies these
observables as arrows of time. Requiring monotonic behaviour of
$N(t)$ for arbitrary values of $\gamma_L, \gamma_S$ and $\Delta m$
yields a bound on the CP-violating overlap  $\delta =
\braket{K_L}{K_S}$, which is similar to, but weaker than, the
known unitarity bound. A similar requirement on $|\vec{\zeta}(t)|$
yields a new bound, $\delta^2 < \frac{1}{2} \left( \frac{\Delta
\gamma}{\Delta m}\right)  \sinh \left( \frac{3\pi}{4} \;
\frac{\Delta \gamma}{\Delta m}\right)$  which is particularly
effective in limiting the CP-violating overlap in the
$B^0$-$\bar{B^0}$ system. We obtain the Stokes parameter $\zeta_3(t)$
which shows how the average strangeness of the beam evolves from
zero to $\delta$. The evolution of the Stokes vector from
$|\vec{\zeta}| = 0$ to $|\vec{\zeta}| = 1$ has a resemblance to an
order parameter of a system undergoing spontaneous symmetry
breaking.

\end{abstract}

\section{Introduction}
We examine in this paper the time evolution of a neutral $K$ meson
prepared as an equal incoherent mixture of $K^0$ and $\bar{K}^0$.
Such a state is easily obtained in a reaction such as $e^+ e^- \to
\phi(1020) \to K^0 \bar{K}^0$, when only one of the kaons in the
final state is observed. (Our considerations apply equally to $B$
mesons produced in $e^+ e^-  \to \Upsilon (4s) \to B^0
\bar{B}^0$.)  An incoherent beam of this type is characterized by
a density matrix, which we write in the $K^0$-$\bar{K}^0$ basis as
\be \rho(t) = \frac{1}{2} N(t) \left[\1 + \vec{\zeta}(t) \cdot
\vec{\sigma} \right] \label{CPV1}\nonumber \ee
The evolution is described
by a normalization function $N(t)$, which is the intensity of the
beam at time $t$, and a Stokes vector $\vec{\zeta}$ which
characterizes the polarization state of the system with respect to
strangeness. The beam, which has $|\vec{\zeta}(0)| =0 $ at the
time of production evolves ultimately into a pure state
corresponding to the long-lived $K$ meson $K_L$, with a Stokes
vector of unit length: $|\vec{\zeta} (\infty)| = 1$. \\ In this
sense, the system can be regarded as possessing two dynamical
functions: $N(t)$ which varies from one to zero, and
$|\vec{\zeta}(t)|$ which goes from zero to one. This evolution
touches on interesting issues such as the role of CP-violation,
and the extent to which the functions $N(t)$ and
$|\vec{\zeta}(t)|$ define arrows of time. The requirement that
these functions are monotonic yields constraints on the
CP-violating parameter $\delta = \braket{K_L}{K_S}$. The fact that
the incoherent initial state is completely neutral with respect to
strangeness and CP quantum numbers is of significance in this
regard. In addition the component $\zeta_3 (t)$ of the Stokes
vector describes the manner in which the strangeness of the state
evolves from zero to final value $\delta = 3.27 \times 10^{-3}$
and serves as a model for flavour-genesis induced by CP violation
in a decaying system.  Finally, the evolution of the system from
an initial ``amorphous'' state with $\vec{\zeta}(0) = 0 $ to a
final ``crystalline'' state described by a three-dimensional
Stokes vector $\vec{\zeta}(t)$ with unit length, is suggestive of
a phase transition, with $\vec{\zeta}(t) $ playing the role of an
order parameter of a system undergoing spontaneous symmetry
breaking.

 \section{Density Matrix}
 An arbitrary state of the $K$ meson can be desccribed by a 2 x 2 density
  matrix which we write, in the $K^0$-$\bar{K}^0$ basis, as~\cite{CVlit1,CVlit2,CVlit3}

 \be
 \rho(t) = \frac{1}{2} N(t) \left[\1 + \vec{\zeta}(t) \cdot \vec{\sigma} \right]
  \label{CPV3}
 \ee
%%%%%%%%%%%%
Here $N(t)$ is the intensity or norm of the state at time $t$,
calculated from the trace of $\rho$,

\be N(t) = \tr \; \rho(t)
 \label{CPV4}
\ee

and $\vec{\zeta}(t)$ is the Stokes vector, whose components can be
expressed as

\be \zeta_i(t) = \tr \left[ \rho(t) \sigma_i \right]  / \tr
\rho(t)
 \label{CPV5}
\ee

An initial state which is a 1:1 incoherent mixture of $K^0$ and $
\bar{K}^0$ has the density matrix

\ba \rho(t) &= \frac{1}{2} \ket{K^0}\bra{K^0} +  \frac{1}{2}
\ket{\bar{K}^0}\bra{\bar{K}^0} \nn &= \frac{1}{2} \left(
\begin{array}{cc}
 1&  0\\
 0& 1
\end{array} \right)
 \label{CPV6}
\end{align}

which corresponds to an initial Stokes vector $\vec{\zeta}(0) =0$.
To determine the time evolution, we note that~\cite{CVlit4}

\ba \ket{K^0} \underset{t}{\to} \ket{\psi(t)} = \frac{1}{2p}
\left[ \ket{K_S} e^{- \lambda_S t} + \ket{K_L} e^{- \lambda_L t}
\right] \nn \ket{\bar{K}^0} \underset{t}{\to}\ket{\bar{\psi}(t)} =
\frac{1}{2q}\left[ \ket{K_S} e^{- \lambda_S t} - \ket{K_L} e^{-
\lambda_L t} \right]
 \label{CPV7}
\end{align}

where we have introduced the eigenstates

\ba \ket{K_L} &= p \ket{K^0} - q \ket{\bar{K}^0} \nn \ket{K_S}&= p
\ket{K^0} +q \ket{\bar{K}^0}    \hspace{2cm}
\raisebox{1em}{$(|p|^2 + |q|^2 = 1)$}
  \label{CPV8}
\end{align}

with eigenvalues

\be \lambda_{L,S} = \frac{1}{2} \gamma_{L,S} + im_{L,S}
 \label{CPV9}
\ee

The overlap of the states $\ket{K_L}$ and $\ket{K_S}$ is given by
the CP-violating parameter

\be \delta = \braket{K_L}{K_S} = (|p|^2 - |q|^2 )/(|p|^2 + |q|^2 )
= 3.27 \times 10^{-3}
 \label{CPV10}
\ee

The resulting density matrix at time $t$ is

\be \rho(t) = \left( \begin{array}{cc}
\rho_{11}(t) & \rho_{12}(t)  \\
\rho_{21}(t)  & \rho_{22}(t)
\end{array} \right)
 \label{CPV11}
\ee

with

\ba \rho_{11}(t) &= \frac{1}{4(1- \delta)} \left[ e^{- \gamma_S t}
+ e^{- \gamma_L t} - 2 \delta e^{- \frac{1}{2} ( \gamma_S +
\gamma_L)t} \cos\Delta mt\right] \nn
\rho_{22}(t) &= \frac{1}{4(1+ \delta)} \left[ e^{- \gamma_S t} +
e^{- \gamma_L t} + 2 \delta e^{- \frac{1}{2} ( \gamma_S +
\gamma_L)t} \cos\Delta mt\right] \nn
\rho_{12}(t) &= \frac{1}{8 p^* q}  \left[ e^{- \gamma_S t} - e^{-
\gamma_L t} + i 2 \delta e^{- \frac{1}{2} ( \gamma_S + \gamma_L)t}
\sin\Delta mt\right] \nn
\rho_{21}(t) &=  \rho_{12}^*(t) \label{CPV12}
\end{align}

Using Eqs. (\ref{CPV4}) and (\ref{CPV5}), we derive

\be N(t) = \frac{1}{2(1-\delta^2)} \left[ e^{-\gamma_S t} + e^{-
\gamma_L t} - 2 \delta^2 e^{ - \frac{1}{2} ( \gamma_S +
\gamma_L)t} \cos\Delta mt \right]\label{CPV13a}
%\nn
\ee
and
\ba\zeta_1(t) &= \frac{2\left[Re \left( p q^*\right) \left(
e^{-\gamma_S t}-e^{- \gamma_L t}\right) - \cI m \left( pq^*\right)
2 \delta e^{ - \frac{1}{2} ( \gamma_S + \gamma_L)t} \cdot
\sin\Delta mt\right]}{e^{-\gamma_S t}+e^{- \gamma_L t}- 2 \delta^2
e^{ - \frac{1}{2} ( \gamma_S + \gamma_L)t} \cos\Delta mt} \nn
\zeta_2(t) &=\frac{-2\left[ \cI m \left(pq^*\right) \left(
e^{-\gamma_S t}-e^{- \gamma_L t}\right)+Re \left( pq^*\right) 2
\delta e^{ - \frac{1}{2} ( \gamma_S + \gamma_L)t} \cdot \sin\Delta
mt\right]}{e^{-\gamma_S t}+e^{- \gamma_L t}- 2 \delta^2 e^{ -
\frac{1}{2} ( \gamma_S + \gamma_L)t} \cos\Delta mt} \nn
\zeta_3(t)& = \delta \frac{e^{-\gamma_S t}+e^{- \gamma_L t}- 2 e^{
- \frac{1}{2} ( \gamma_S + \gamma_L)t} \cos\Delta mt}{e^{-\gamma_S
t}+e^{- \gamma_L t}- 2 \delta^2 e^{ - \frac{1}{2} ( \gamma_S +
\gamma_L)t} \cos \Delta mt} \label{CPV13}\enspace .
\end{align}

Note that the components $\zeta_{1,2}(t)$ involve $Re \left(
pq^*\right) $ and $\cI m \left( pq^*\right) $ where $p$ and $q$
are the coefficients in the definition of $K_{L,S}$ in eq.(\ref{CPV8}).
 These are convention-dependent, since the relative
phase of $p$ and $q$ can be changed by a phase transformation
$\ket{K^0} \to e^{i \alpha} \ket{K^0}$,
$\ket{\bar{K}^0} \to e^{-i\alpha} \ket{\bar{K}^0}$. A quantity independent of phase
convention is

\be
\zeta_1^2+\zeta_2^2 = (1 - \delta^2)  \frac{\left( e^{-\gamma_S t}-e^{- \gamma_L t}\right)^2
 + 4  \delta^2 e^{ -\left( \gamma_S +\gamma_L \right) t} \sin^2\Delta mt}
 { \left[ e^{-\gamma_S t}
 +e^{-\gamma_L t} -2 \delta^2 e^{ - \frac{1}{2} ( \gamma_S + \gamma_L)t} \cos\Delta mt \right]^2}
\label{CPV14}
\ee

Thus the length of the Stokes vector is

\ba
|\vec{\zeta}(t)| &= \left[ \zeta_1^2(t) + \zeta_2^2(t) + \zeta_3^2(t) \right]^{\frac{1}{2}}\nn
&= \frac{1}{2 N(t) (1-\delta^2)} \left[ \delta^2\left\lbrace e^{-\gamma_S t}+e^{-\gamma_L t}
 - 2 e^{ - \frac{1}{2} ( \gamma_S + \gamma_L)t} \cos\Delta mt\right\rbrace^2 \right. \nn
 &+\left. (1-\delta^2)\left\lbrace \left( e^{-\gamma_S t} - e^{-\gamma_L t}\right)^2  +
 4  \delta^2 e^{ - \frac{1}{2} ( \gamma_S + \gamma_L)t} \cdot \sin^2 \Delta mt
  \right\rbrace \right] \nn
&= \left[1- \frac{1}{N(t)^2} e^{- ( \gamma_S + \gamma_L)t}\right]^{\frac{1}{2}}
\label{CPV15}
\end{align}

This equation provides a simple relation between the magnitude of
the Stokes vector $|\vec{\zeta}(t)|$ and the normalization
function $N(t)$.

In the CP-invariant limit, $\delta \to 0$, the density matrix reduces to

\be
\rho(t) \underset{\delta \to 0}{\longrightarrow} \frac{1}{4} \left[ \begin{array}{cc}
 e^{- \gamma_S t} + e^{- \gamma_L t}&  e^{- \gamma_S t} - e^{- \gamma_L t} \\
 e^{- \gamma_S t} - e^{- \gamma_L t} &  e^{- \gamma_S t} + e^{- \gamma_L t}
\end{array} \right]
\label{CPV16}
\ee

and the limiting form of $N(t)$ and $\vec{\zeta}(t)$ is

\ba
N(t) & \longrightarrow \frac{1}{2}  \left[ e^{- \gamma_S t} + e^{- \gamma_L t}\right]  \nn
\zeta_{12} \equiv \left[ \zeta^2_1 (t) + \zeta_2^2(t)\right] ^{\frac{1}{2}} &  \longrightarrow
 \left( e^{- \gamma_L t} - e^{- \gamma_S t}\right) /\left( e^{- \gamma_S t} + e^{- \gamma_L t}\right)
\nn
\zeta_3(t) & \longrightarrow 0
\label{CPV17}
\end{align}

\begin{figure}
\begin{center}
\includegraphics[width=0.98\textwidth]{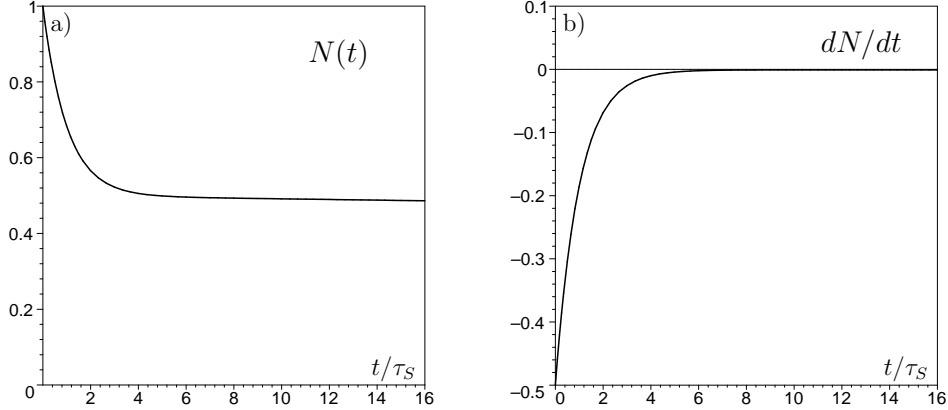}
%{\small A. A. Friedmann 1888-1925}
\end{center}
\caption{\small{Normalization function $N(t)$ (a) and its time derivative (b) as function of time
in units of $\tau_S$ for a state prepared
as incoherent mixture of $K^0$ and $\bar{K}^0$.}}
\label{CPVfig1}
\end{figure}

The behaviour of $N(t)$ and $dN/dt$ for the $K^0$-$\bar{K}^0$
system is shown in fig.\ref{CPVfig1}.

%\end{minipage}
%\begin{minipage}{0.45\textwidth}
%\end{minipage}

\begin{figure}
\begin{center}
\includegraphics[width=0.99\textwidth]{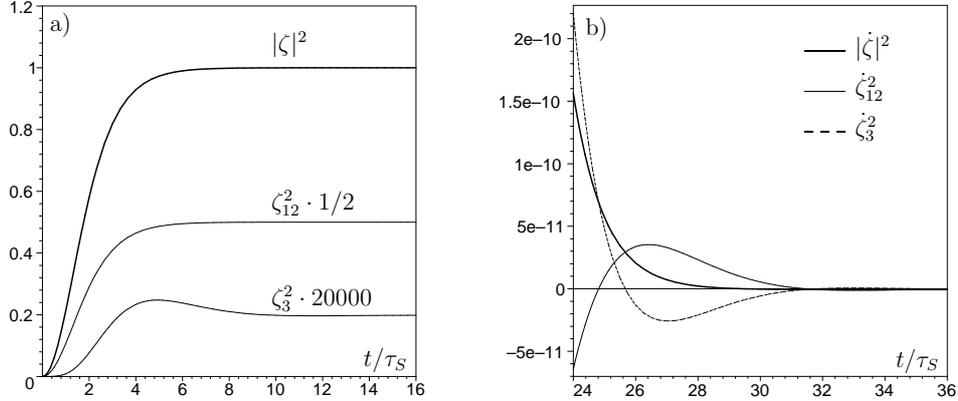}
%{\small A. A. Friedmann 1888-1925}
\end{center}
\caption{{\small Behaviour of $|\zeta|^2$, $\zeta_{12}^2$ and $\zeta_3^2$
(a) and their time derivatives (b) as function of time.
Note the different scale for  $\zeta_{12}^2$ and $\zeta_3^2$ in (a). In (b) only the tail
of the time dependence is shown. }}
\label{CPVfig2}\end{figure}

The behaviour of the functions $\zeta_{12}^2 (t), \zeta_3^2 (t)$
and $\zeta^2(t) = \zeta_{12}^2 (t)+ \zeta_3^2(t)$ and their time
derivatives is shown in fig.\ref{CPVfig2}. The function
$\zeta_3^2$ is clearly nonmonotonic, and its derivative has a
number of zeros (e.g. $t/\tau_S= 4.95,11.6,18.2,...$). By comparison
the derivative of $\zeta_{12}^2$ has a distant zero at
$t/\tau_S=25.7$. As seen in fig.\ref{CPVfig2}b these two
nonmonotonic functions combine to produce a Stokes vector
$\zeta^2(t)$ which is strictly monotonic, the asymptotic values
being $\zeta_{12}^2\rightarrow (1- \delta^2)$, $\zeta_3^2\rightarrow
\delta^2$, $\zeta^2\rightarrow 1$.

%\end{minipage}
%\begin{minipage}{0.45\textwidth}
%\end{minipage}

\section{Arrows of time}
\subsection{The Normalization Arrow $N(t)$}

The normalization of the kaon state is given in eq.(\ref{CPV13a}).
% and reads
%\be
%N(t) = \frac{1}{2(1-\delta^2)} \left[ e^{- \gamma_S t} + e^{- \gamma_L t}
% - 2 \delta^2 e^{ - \frac{1}{2} ( \gamma_S + \gamma_L)t} \cos \Delta mt \right]
%\label{CPV18}
%\ee
As seen in fig.\ref{CPVfig1}, this function is indeed monotonic for the parameters
of the $K$ meson system.
This monotonic (unidirectional) property implies
that $N(t)$ behaves as an arrow of time. In the absence of CP-violation
($\delta=0$), the function $N(t)$ is simply the sum of two exponentials
$(e^{- \gamma_S t} + e^{- \gamma_L t})/2$,
and the monotonic decrease is ensured by the requirement
$\gamma_S, \gamma_L >0$ (positivity of the decay matrix).
The third term in eq.(\ref{CPV13a}),
appearing when $\delta \neq 0$, indicates a $K_L$-$K_S$ interference effect.
It implies that an incoherent $K^0$-$\bar{K}^0$ mixture does
not evolve like an incoherent $K_L$-$K_S$ mixture.
Notice however, that the coefficient of the interference term is
quadratic in $\delta$, so that the function $N(t)$ is
CP-even, remaining unchanged under $\delta \to - \delta$.
Nevertheless the presence of the $\delta^2$ term is decisive in determining
whether or not $N(t)$ is monotonic, and hence an arrow of time.
If we require the function $N(t)$ to be monotonic ($dN/dt <0$)
then we have from (\ref{CPV13a}) (see also~\cite{CVlit5}) that
%
%The requirement of monotonicity, $d N/dt < 0$, yields~\cite{CVlit5}
%
\ba
\frac{dN}{dt} = \frac{-1}{2(1-\delta^2)} \left[ \gamma_S e^{- \gamma_S t}
 + \gamma_L e^{-\gamma_L t} - 2 \delta^2 e^{-\frac{1}{2}\left( \gamma_S + \gamma_L \right) t}\right.  \nn
\left. \cdot \left\lbrace \frac{\gamma_S + \gamma_L}{2} \cos \Delta m t +
  \Delta m \sin \Delta m t \right\rbrace \right]  < 0
\label{CPV19}
\end{align}

from which it follows, as sufficient condition, that

\ba
&\delta^2 \leq  \left( \frac{\gamma_S \gamma_L}{
  \left( \gamma_S+ \gamma_L\right)^2\!\!/4 + \Delta m^2}\right)^{\frac{1}{2}} \nn
\mbox{or} \qquad & \delta^2 \leq  \left(\frac{r}{
\left(1+ r\right)^2\!\!/4 +\mu^2}\right)^\frac{1}{2}
\label{CPV20}
\end{align}

where we have introduced the notation $r= \gamma_L /  \gamma_S,
\mu = \Delta m /  \gamma_S$.  This constraint is analogous to, but
weaker than, the unitarity constraint derived in~\cite{CVlit6,CVlit7}, which
reads

\be
\delta^2_{{\rm unit}} \leq \frac{r}{
\left(1+ r\right)^2\!\!/4 +\mu^2} \enspace .
\label{CPV28}
\ee

It is clear that the unitarity bound is interesting for a system
like $K^0$-$\bar{K}^0$, where $r = \gamma_L/ \gamma_S$ is small,
but does not provide a useful constraint for $B^0$-$\bar{B}^0$,
where $r$ is close to 1.

\begin{figure}
\begin{center}
\includegraphics[width=0.98\textwidth]{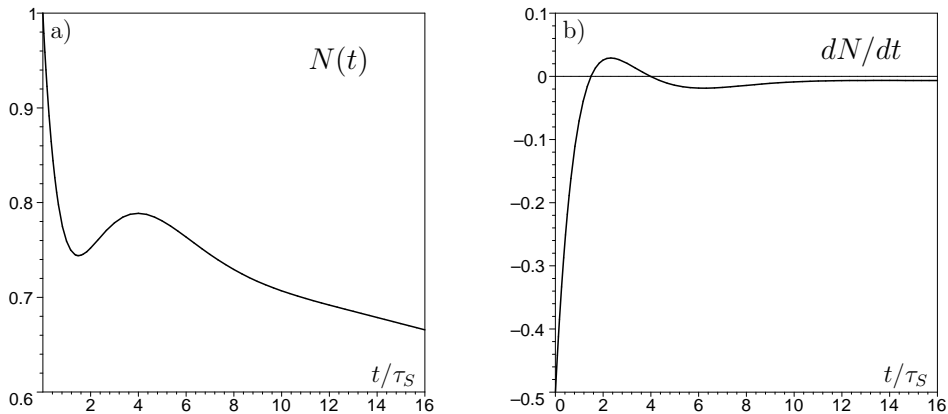}
%{\small A. A. Friedmann 1888-1925}
\end{center}
\caption{{\small Behaviour of $N(t)$ (a) and $dN/dt$ (b) for values of parameters
$\mu=\mu_K,\,\, r=0.01,\,\, \delta= 0.6$ violating the
monotonicity bound (\ref{CPV20}).}}
\label{CPVfig3}\end{figure}

To see what happens if the parameters $\delta, r$ and $\mu$ are allowed to vary,
we show in fig.\ref{CPVfig3} the behaviour of $N(t)$ and $dN/dt$ for $\delta =0.6,
\,\, r =0.01 $,
keeping $\mu$ at its standard $K$ meson value, $\mu_K = 0.47$.
The function $N(t)$ shows fluctuations, and the derivative $dN/dt$ changes sign.
Such a behaviour results from the violation of the bound (\ref{CPV20}).
The fluctuations in $N(t)$ may be regarded as fluctuations in the direction of the time arrow
(we call this phenomenon ``Zeitzitter''), and can occur when the CP-violating parameter
$\delta^2$ exceeds the limit (\ref{CPV20}).  This is the manner in which CP violation impacts
on the time arrow, even though the function $N(t)$ is CP-even.

%\end{minipage}
%\begin{minipage}{0.45\textwidth}
%\end{minipage}

From the point of view of an observer monitoring the intensity of
the kaon beam (for example by measuring the rate of leptonic
decays $\pi^{\mp} \ell^{\pm}\nu$) the fluctuation in $N$ would appear as
an inexplicable enhancement or suppression of the beam intensity
in certain intervals of time. The effect can be regarded
equivalently as violation of unitarity or a flutter in the arrow
of time.

\subsection{The Stokes Arrow}
%\end{minipage}
%\begin{minipage}{0.45\textwidth}
%\end{minipage}
The magnitude of the Stokes vector $|\vec{\zeta}(t)|^2$ calculated
in eq.(\ref{CPV15}), is a measure of the coherence of the state,
and is plotted in fig.\ref{CPVfig2}
%and fig.\ref{CPVfig4}a
for the physical $K$-meson
parameters. One sees that the function $|\vec{\zeta}|^2$ evolves
monotonically from 0 to 1, and its derivative remains positive at
all times. Thus the Stokes parameter $|\vec{\zeta}(t)|$ qualifies
as an arrow of time. To see how this arrow is affected if the
parameters $\delta, r$ and $\mu$ are allowed to vary, we look at
the derivative of the function $\zeta(t)$. Writing

\be
|\vec{\zeta}(t)| = \left[ 1 - \frac{e^{-\left( \gamma_S + \gamma_L\right) t}}
{N(t)^2}\right]^{\frac{1}{2}}
\label{CPV21}
\ee

we find that the monotonicity condition $d |\vec{\zeta}(t)| /dt >
0$ is equivalent to the condition

\be
\left( \frac{d N}{dt} + \frac{1}{2} \left( \gamma_S + \gamma_L\right) N\right) \leq 0
\label{CPV22}
\ee

which implies

\be
\left( e^{\Delta \gamma t/2} - e^{- \Delta \gamma t/2}\right)
  + 4 \delta^2 \frac{\Delta m}{\Delta \gamma} \sin \Delta m t \geq 0
\label{CPV23}
\ee

where $ \Delta \gamma = \gamma_S - \gamma_L$. From this we  derive
a new upper bound on $\delta^2$:

\ba
& \delta^2 < \frac{1}{2} \left( \frac{ \Delta \gamma }{\Delta m}\right)
  \sinh \left( \frac{3\pi}{4} \; \frac{ \Delta \gamma }{\Delta m}\right)   \nn
\mbox{or} \qquad &\delta^2 < \frac{1}{2} \left( \frac{1- r}{\mu}\right)
 \sinh \left( \frac{3\pi}{4} \; \frac{\left( 1- r\right) }{\mu}\right)
\label{CPV24}
\end{align}

\begin{figure}[h]
\begin{center}
\includegraphics[width=0.8\textwidth]{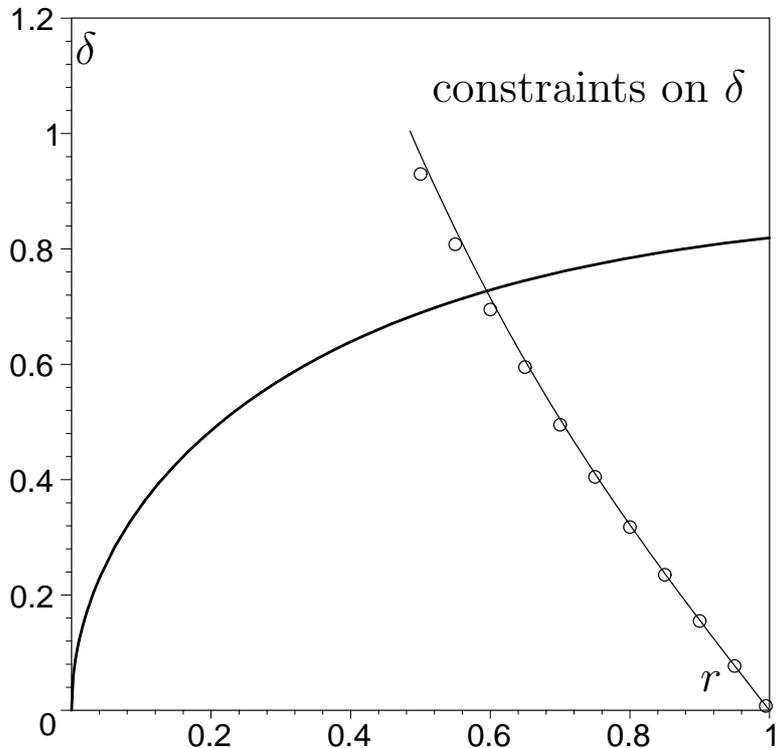}
%{\small A. A. Friedmann 1888-1925}
\end{center}
\caption{{\small Constraints on $\delta$ in the $\delta$-$r$ plane
resulting from unitarity and monotoni\-city
of $|\zeta(t)|$ for a $B^0$-like system with $\mu=0.7$. The
thick line represents the unitarity bound (\ref{CPV28}) and the thin line our new bound
evaluated from (\ref{CPV24}). The numerical evaluation of (\ref{CPV23}) (open circles)
yields values very close  to the approximation given in eq.(\ref{CPV24}). }}
\label{CPVfig5}\end{figure}

This bound is obtained from the requirement that
$|\vec{\zeta}(t)|$ be monotonic (an arrow of time) just as the bound
in eq.(\ref{CPV20}) was derived from the monotonicity of $N(t)$.
The bound (\ref{CPV24}) is particularly effective in constraining the value of the overlap
parameter in the $B^0$-$\bar{B}^0$ system, in which the decay widths of
the two eigenstates are close together, $r \to 1$.
In this respect the bound in eq.(\ref{CPV24}) is complementary to the unitarity
bound in eq.(\ref{CPV28}) which is effective when $r \to 0$.
The contrast between the two bounds is highlighted in fig.\ref{CPVfig5}.
Taking the parameters of the $B^0$-$\bar{B}^0$ system to be $ r = 0.99, \mu = 0.7$,
we obtain from (\ref{CPV24})

\be
\delta_B = \braket{B^0_S}{B^0_L} \lesssim 0.0155
\label{CPV25}\enspace .
\ee

\begin{figure}[h]
\begin{center}
\includegraphics[width=0.78\textwidth]{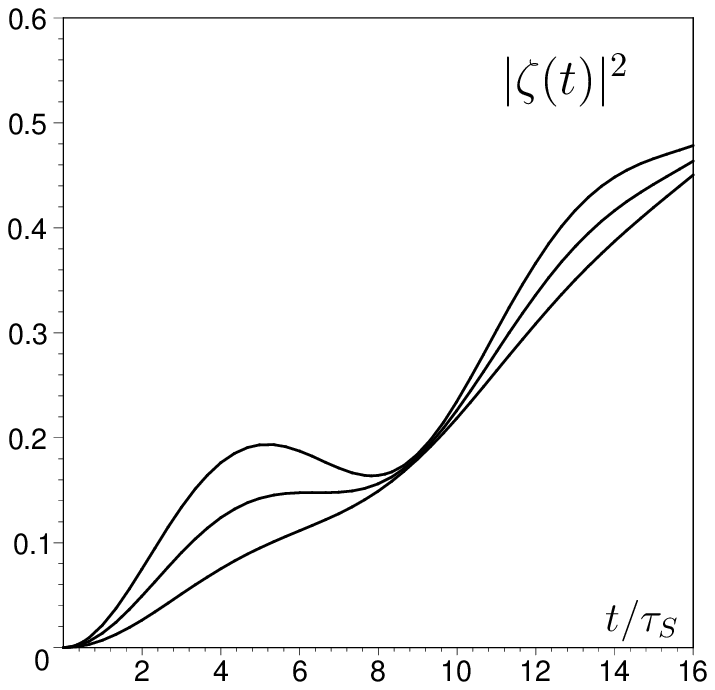}
%{\small A. A. Friedmann 1888-1925}
\end{center}
\caption{{\small Evolution of the Stokes vector $|\zeta(t)|^2$
for a $B^0$-like system with para\-meters $\mu_B=0.7,\,\, r=0.9$.
The middle curve corresponds to $\delta_{\rm crit}=0.156$ obtained from the bound (\ref{CPV24}).
The nonmonotonic upper curve is obtained  for   $\delta=0.2$
and the monotonic lower curve for  $\delta=0.1$.}}
\label{CPVfig4}\end{figure}

%If the parameters of the $B$-system are chosen so as to violate this bound,
%the time dependence of $ |\vec{\zeta}(t)|$ develops fluctuations (fig. \ref{CPVfig4}).
We wish to stress that for a $B^0$-like system the bound in
(\ref{CPV24}) is not just a sufficient condition for monotonic
behaviour of $|\zeta(t)|^2$, but almost a critical value
separating the monotonic and nonmonotonic domains. As an
illustration we show in fig.\ref{CPVfig4} the transition in the
behaviour of $|\zeta(t)|^2$ for a system with parameters
$\mu=0.7,\,\, r=0.9$, as $\delta$ is varied from a value 0.1, below
the critical value of $\delta_{\rm crit}=0.156$, to a value $0.2$
above $\delta_{\rm crit}$.

The fluctuations in $|\zeta(t)|^2$, shown in fig.\ref{CPVfig4} are the analog
of the fluctuations in $N(t)$, shown in fig.\ref{CPVfig3},
which arise when the parameters of the system violate the bound in eq.(\ref{CPV20}).
Whereas the fluctuation in $N(t)$ would reveal itself as an inexplicable Zitter in the
beam intensity, the fluctuation in  $ |\vec{\zeta}(t)|$  would show up as an unaccountable
Zitter in the coherence of the beam.
In both cases, the effect results from a breakdown in the monotonicity of a function,
associated with a loss of directionality in an arrow of time.\\

\section{Evolution of Strangeness}

The component  $\zeta_3(t)$ of the Stokes vector has a special
 significance: it is the expectation value of $\sigma_3$, which can be identified with
  the strangeness operator with eigenvalues +1 for $K^0$ and $-1$ for $\bar{K}^0$.
Thus a measurement of $\zeta_3(t)$ is simply a measurement of the decay asymmetry
  into the channels $ \pi^- \ell^+\nu$ and $\pi^+ \ell^- \bar\nu$ :

\be
\zeta_3(t) = \frac{\Gamma \left( \pi^- \ell^+ \nu;t\right) -
\Gamma \left( \pi^+ \ell^-\bar\nu; t\right) }{\Gamma \left( \pi^- \ell^+ \nu;t\right)
 +\Gamma \left( \pi^+ \ell^- \bar\nu; t\right)}
\label{CPV26}
\ee

\begin{figure}
\begin{center}
\includegraphics[width=0.98\textwidth]{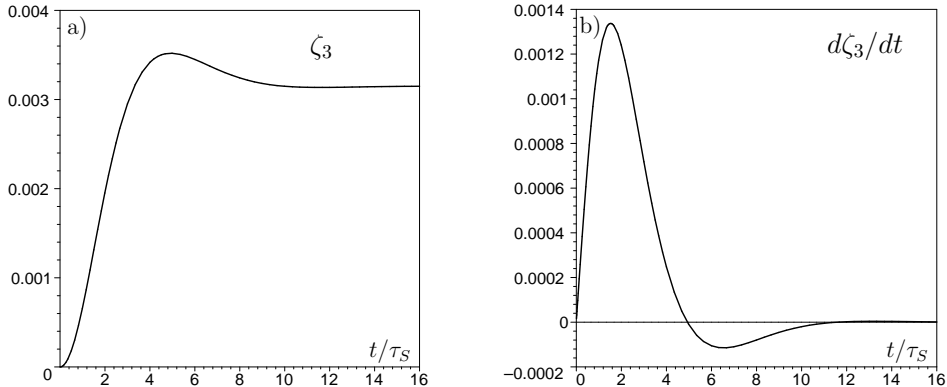}
%{\small A. A. Friedmann 1888-1925}
\end{center}
\caption{{\small Evolution of $\zeta_3(t)$  (a) and its derivative (b)
in the $K^0$-$\bar{K}^0$ system.}}
\label{CPVfig6}\end{figure}

Referring to eq.(\ref{CPV13}), we observe that $\zeta_3(t)$ is a pure CP-violating
observable, since it changes sign under $\delta \to - \delta$
(By contrast, the functions $N(t)$ and $|\vec{\zeta}(t)|$, are invariant
 under $\delta \to - \delta$). Writing $\zeta_3(t)$ explicitly as

\be
\zeta_3(t) =\delta\frac{e^{-\gamma_S t} + e^{- \gamma_L t}
 - 2 e^{- \frac{1}{2}\left( \gamma_L + \gamma_S\right) t} \cos \Delta m
 t}{ 2\left( 1-\delta^2\right)  N(t)}
\label{CPV27}
\ee

we note that it is a quotient of
a function that contains an oscillating term and a monotonic function $N(t)$.
The average strangeness $\zeta_3(t)$
is thus clearly not a monotonic function of time. This is visible in fig.\ref{CPVfig6},
where we also show the derivative $d \zeta_3 (t)/dt$.
We have here an explicit example of a CP-odd observable emerging from an initial
state that has no preferred CP direction. Such observables are not monotonic,
and cannot be associated with an arrow of time.

%\end{minipage}
%\begin{minipage}{0.45\textwidth}
%\end{minipage}

\section{Summary}
\begin{itemize}
\item[(1)] We have shown that the evolution of an incoherent
 $K^0$-$\bar{K}^0$ mixture is characterized by two time-dependent
functions, the norm $N(t)$ and the magnitude of the Stokes vector
$|\vec{\zeta}(t)|$ both of which evolve monotonically and may
therefore be associated with microscopic arrows of time. It should
be stressed that we are discussing here \emph{conditional}
arrows of time, whose existence depends on the degree of CP violation, and
not simply on the positivity of the decay widths $\gamma_{L,S}$.

\item[(2)]  If the parameters $\gamma_L, \gamma_S, \Delta m $ and
$ \delta= \braket{K_L}{K_S}$ are allowed to vary, the requirement
of monotonic behaviour of $N(t)$ leads to the bound in eq.(\ref{CPV20}),
 which is similar to, but weaker than, the unitarity
bound (\ref{CPV28}), derived in~\cite{CVlit6,CVlit7}.  The requirement of
monotonicity for $|\vec{\zeta}(t)|$ leads to a new bound on
$\delta^2$  given in eq.(\ref{CPV24}), which is complementary to
the unitarity bound (\ref{CPV28}), and far more restrictive for
systems such as $B^0$-$\bar{B}^0$  with  $r = \gamma_L/ \gamma_S
$ close to unity.

\item[(3)] A violation of the bounds in eq.(\ref{CPV20}) and
(\ref{CPV24}) leads to fluctuations in $N(t)$ and $\zeta(t)$
associated with fluctuations in the arrow of time (``Zeitzitter'')
and a violation of unitarity.

\item[(4)] It is worth noting that the product $N^2 (1-
|\vec{\zeta}|^2)$ is equal to $e^{-(\gamma_L +
\gamma_S)t}$ and therefore monotonic for all values of $\delta,r$ and $\mu$.
This product is just four times the determinant of the density matrix
$\rho(t)$,

\item[(5)] The time-dependence of $\zeta_3(t)$ describes the
evolution of strangeness in a beam that is initially an equal
mixture of $K^0$ and $\bar{K}^0$. It is an example of
flavour-genesis induced by CP violation in a decaying system.

\item[(6)] The emergence of a non-zero three-dimensional Stokes
vector $\vec{\zeta}(t)$ from a state that is initially
``amorphous'' $( \vec{\zeta}(0) =0)$, is suggestive of a phase
transition. The evolution of the Stokes vector from zero to unit
length is reminiscent of an order parameter for a system
undergoing spontaneous symmetry breaking.

\item[(7)] All our considerations have been in the framework of
ordinary quantum mechanics and CPT invariance. Discussions that
involve violation of quantum mechanics and/or CPT symmetry may be
found, for example, in~\cite{CVlit8}. An early discussion of the arrow
of time in connection with $K$ meson decays is given in~\cite{CVlit9}. Broader
issues connected with the arrow of time are discussed, for
instance, in~\cite{CVlit10}. Finally, experimental investigations of
discrete symmetries in the decays of $K$ mesons and $B$ mesons produced
in $e^+ e^-$ or $p \bar{p}$ collisions are described in~\cite{CVlit11}.
\end{itemize}

\paragraph{Acknowledgement:} One of us (LMS) wishes to thank
 Dagmar Bruss (University of D{\"u}sseldorf)
 for a useful correspondence.

\vskip3em

%{\Large\bf{References}}\\


\begin{thebibliography}{xx}
\bibitem{CVlit1} U. Fano, Rev. Mod. Phys. \underline{29}, 74 (1957)
\bibitem{CVlit2} R. G. Sachs, \emph{Physics of Time Reversal}, University of Chicago Press, 1987
\bibitem{CVlit3} L. M. Sehgal \emph{Density Matrix Description of Neutral K Meson Decay},
 Tata Institute report, TIFR - TH- 70-35 (1970)
\bibitem{CVlit4} T. D. Lee, R. Oehme and C. N. Yang, Phys. Rev. \underline{105}, 1671 (1957)
\bibitem{CVlit5} L. M. Sehgal, \emph{Decays of Neutral K Mesons
Produced in $ p \bar{p}$ Annihilation. A Comment}, Aachen preprint, 1973 (unpublished)
\bibitem{CVlit6} T. D. Lee and L. Wolfenstein, Phys. Rev. 138, B1490 (1965)
\bibitem{CVlit7} J. S. Bell and J. Steinberger,
in Proc. Oxford International Conference on Elementary Particles, 1965, pp. 195-222
\bibitem{CVlit8} P. Huet, M. E. Peskin, Nucl. Phys. B 434, 3 (1995)\\
J. Ellis, N. E. Mavromatos and D. V. Nanopoulos, Phys. Lett. B 293, 142 (1992)
\bibitem{CVlit9} A Aharony, Ann. Phys. \underline{67}, 1 (1971); ibid \underline{68}, 163
(1971);\\
A. Aharony and Y. Ne'eman, Int. J. Theor. Phys. \underline{3}, 437 (1970)
\bibitem{CVlit10} \emph{Physical Origins of Time Asymmetry},
 Eds. Jose Angel Sanchez Asiain, et al, Cambridge University Press (1996)
\bibitem{CVlit11} KLOE Collaboration, F. Ambrosino et al, Phys.Lett.
\underline{B642}, 315 (2006)\\
Babar Collaboration (B.Aubert et al), Phys.Rev.Lett. \underline{96}, 251802
(2006)\\
Belle Collaboration (E.Nakano et al), Phys.Rev. \underline{D73}, 112002 (2006)
CPLEAR Collaboration, M. Carrol et al, Nucl. Phys. A 626 157c-165c (1997); \\
K. Kleinknecht \emph{Uncovering CP Violation}, Springer-Verlag, Berlin, Heidelberg (2003).
\end{thebibliography}
\end{document}